\documentclass[aps,pra,twocolumn,showpacs,superscriptaddress]{revtex4-1}
\usepackage{units}
\usepackage{amsmath}
\usepackage{amssymb}
\usepackage{graphicx}
\usepackage{bm}
\usepackage{multirow,color,ulem}

\newcommand{\hb}{}
\newcommand{\re}{\text{Re}}
\newcommand{\im}{\text{Im}}
\newcommand{\gper}{\gamma_\perp}
\newcommand{\gpar}{\gamma_\parallel}

\newcommand{\be}{\begin{equation}}
\newcommand{\ee}{\end{equation}}
\newcommand{\bea}{\begin{eqnarray}}
\newcommand{\eea}{\end{eqnarray}}
\newcommand{\oma}{\omega_{a}}
\newcommand{\oml}{\omega_{L}}
\newcommand{\omc}{\omega_{r}}
\newcommand{\Oms}{\Omega_{s}}
\newcommand{\omM}{\omega_m}
\newcommand{\DLa}{\Delta_{La}}
\newcommand{\DLr}{\Delta_{Lr}}
\newcommand{\tDLr}{\tilde{\Delta}_{Lr}}
\newcommand{\gmu}{\kappa}

\newcommand{\gM}{\Gamma_m}
\newcommand{\epso}{\epsilon_{0}}
\newcommand{\dn}{\delta\hat{n}}
\newcommand{\dD}{\delta\hat{D}}

\newcommand{\F}{\hat{\mathcal{F}}}
\newcommand{\Fa}{\F_{a}}
\newcommand{\Fd}{\F_{\gpar}}
\newcommand{\nsat}{n_{sat}}

\newcommand{\Daad}{\mathcal{D}_{aa^\dagger}}
\newcommand{\Dada}{\mathcal{D}_{a^\dagger a}}
\newcommand{\Dadd}{\mathcal{D}_{a^\dagger \gpar}}

\newcommand{\Dad}{\mathcal{D}_{a\gpar}}

\newcommand{\Ddd}{\mathcal{D}_{\gpar\gpar}}
\newcommand{\Gopt}{\Gamma_\text{opt}}

\newcommand{\avg}[1]{\langle{#1}\rangle}

\newcommand{\Eq}[1]{Eq.~(\ref{#1})}

\newcommand{\Fig}[1]{Fig.~\ref{#1}}

\begin{document}

\title{Gain-tunable optomechanical cooling in a laser cavity}

\author{Li Ge}
\affiliation{Department of Electrical Engineering, Princeton University, Princeton, New Jersey 08544, USA}
\author{Sanli Faez}
\affiliation{Max-Planck-Institute for the Science of Light, G\"unther-Scharowsky-Stra{\ss}e 1/Bau 24, DE-91058 Erlangen, Germany}
\author{Florian Marquardt}
\affiliation{Max-Planck-Institute for the Science of Light, G\"unther-Scharowsky-Stra{\ss}e 1/Bau 24, DE-91058 Erlangen, Germany}
\affiliation{Institute for Theoretical Physics, Universit\"at Erlangen-N\"urnberg, Staudtstra{\ss}e 7, DE-91058 Erlangen, Germany}
\author{Hakan~E.~T\"ureci}
\affiliation{Department of Electrical Engineering, Princeton University, Princeton, New Jersey 08544, USA}

\date{\today}

\begin{abstract}
We study the optical cooling of the resonator mirror in a cavity-optomechanical system that contains an optical gain medium. We find that the optical damping rate is vanishingly small for an incoherently pumped laser above threshold. In the presence of an additional external coherent drive however, the optical damping rate can be enhanced substantially with respect to that of a passive cavity. We show that the strength of the incoherent pump provides the means to tune the optical damping rate and the steady state phonon number. The system is found to undergo a transition from the weak optomechanical coupling regime to the strong optomechanical coupling regime as the strength of the incoherent pump is varied.
\end{abstract}


\maketitle

Cavity optomechanics is a rapidly growing field of research studying the coupling of mechanical degrees of freedom to modes of optical cavities \cite{Kippenberg08,Marquardt09}. One fascinating consequence of this interaction is the possibility to laser-cool the mechanical motion (e.g. \cite{Gigan06,Arcizet06,Schliesser,Thompson}). Theory predicted ground state cooling to be possible \cite{Marquardt07,WilsonRae07,Genes08}, and recent experiments \cite{Teufel,Chan} reached down to less than one phonon.
Most of the earlier work in this area has focused on optomechanics in a passive cavity setup with mechanical elements, with relatively little attention on non-standard setups including optical non-linearities \cite{Agarwal}.

In this Letter we address the following question: can a laser cool its own mirrors? More specifically, we consider a cavity-optomechanical system with an incoherently pumped optical gain medium, modeled as an ensemble of identical two-level systems that couples to the cavity photon field. We find that the answer is surprisingly not affirmative. While a laser can cool a second (passive) cavity, as studied extensively in earlier works, it is very inefficient in cooling its own mirrors. This is attributed to the saturating non-linearity that gives rise to a vanishing restoring force on cavity field fluctuations. In the presence of an additional coherent ``seeding" signal however, we find that the damping rate can be significantly enhanced with respect to that of a passive cavity.  We attribute this effect to a reduced effective cavity decay rate that can be tuned by the strength of the incoherent pump. We subsequently discuss the minimum attainable phonon number and show that it can be lowered from that of a passive cavity if the thermal contribution is non-negligible. In the end we discuss the phonon spectrum and show a transition from the weak optomechanical coupling regime to the strong optomechanical coupling regime as the incoherent pump strength is varied.

We first consider the case without the seeding signal. The Hamiltonian can be written as
\begin{align}
\hat{H} &= \hb\omc(1-\frac{\hat{x}}{L})(\hat{a}^\dagger\hat{a}-\bar{n}) + \hb\omM\hat{b}^\dagger\hat{b} + \frac{\hb\oma}{2}\sum_j\hat{\sigma}^z_{j} \nonumber \\
&\quad   + g\sum_j \hb (\hat{a}^\dagger\hat{\sigma}^{-}_{j} + \hat{\sigma}^+_j\hat{a}) + \hat{H}_\text{bath} + \hat{H}_\text{pump}, \label{eq:H}
\end{align}
in which we have set $\hbar=1$. $\hat{a}^\dagger, \hat{a}$ and $\hat{b}^\dagger, \hat{b}$ are the creation and annihilation operators of the cavity photon field of frequency $\omc$ and the mechanical eigenmode of frequency $\omM$, respectively. $L$ is the length of the optical cavity, $\bar{n}$ is the average cavity photon number in the steady state, and $\hat{x}\equiv x_\text{ZPF}(\hat{b}+\hat{b}^\dagger)$ in which the zero point fluctuation of the mechanical oscillator $x_\text{ZPF} = (2m\omM)^{-\frac{1}{2}}$.
The first two terms in (\ref{eq:H}) describe the corresponding passive optomechanical system \cite{Marquardt07}; the next two describe the active medium of transition frequency $\oma$ and its coupling to the cavity field with strength $g$. We have employed the rotating wave approximation, and the subscript $j$ indicates each two-level system in the ensemble. Below we use the collective operators $\hat{P}\equiv\sum_j\sigma^+_j,\,\hat{D}\equiv\sum_j\sigma^z_j$ whose expectation values give the macroscopic polarization and inversion of the medium. $\hat{H}_\text{pump}$ describes the incoherent pump and the decay of the 2-level system it introduces, and $\hat{H}_\text{bath}$ represents all the other decay processes.


A first insight into the modification of opto-mechanical interaction due to the atomic medium can be obtained by considering the optical damping rate $\Gopt$ from Fermi's golden rule in the rate equation approach \cite{Marquardt07}:
\be
\Gopt(\omM) = G^2 [S_{nn}(\omM)-S_{nn}(-\omM)]. \label{eq:GammaOpt}
\ee
Here $G\equiv \omc x_{\text{ZPF}}/{L}$ is the optomechanical coupling strength, and $S_{nn}(\omega)$ is the autocorrelation function of the cavity photon number, defined as the Fourier transform ($F(\omega) \equiv \int_{-\infty}^\infty F(t) e^{i\omega t} dt$) of
\be
S_{nn}(t) = \langle\hat{a}^\dagger(t)\hat{a}(t)\hat{a}^\dagger(0)\hat{a}(0)\rangle - \langle \hat{a}^\dagger(t)\hat{a}(t)\rangle^2. \label{eq:Snn0}
\ee
$\Gopt(\omM)$ defined above applies to both the Hamiltonian (\ref{eq:H}) and a passive system in which no gain medium is present. The coupling between the cavity field and the gain medium brings the system into a self-organized optical oscillation in the presence of the incoherent pump, centered at the laser frequency $\oml$. Consequently, the relaxation dynamics of $S_{nn}(t)$ is modified with respect to that of a passive system and may display multi-mode or chaotic dynamics in general.

Assuming the decay of the polarization ($\gper$) is much faster than that of inversion ($\gpar$) as in a Class A or B laser, we first adiabatically eliminate $\hat{P}$ in the rotating frame of the laser frequency $\oml$ (see Appendix \ref{sec:EOM}). By linearizing the operators about their classical steady-state values, i.e. $\hat{a}(t) = (\bar{a} + \delta\hat{a}(t))e^{-i\oml t},\,\hat{D}(t) = \bar{D} + \delta\hat{D}(t)$, we obtain the following equation of motion
\begin{align}
\dot{\delta\hat{a}}(t) &= \frac{g^2}{\gper-i\DLa}\bar{a} \delta\hat{D}(t) + \Fa(t), \label{eq:a1}
\end{align}
above the laser threshold and in the absence of the optomechanical coupling. Here $|\bar{a}|^2=\bar{n}$, $\DLa \equiv \oml - \oma$ is the detuning from the atomic transition frequency, and $\Fa$ is the effective fluctuation force on $\delta\hat{a}$ (see Appendix \ref{sec:EOM}). The incoherent pump strength is quantified by the steady-state value of the inversion $\bar{D}$ in the absence of the cavity field, denoted by $D_0$.  Above the laser threshold, i.e. $D_0>D_{th}={2\gmu}/{W}$, gain saturation clamps $\bar{D}$ to $D_{th}$ \cite{Haken,LaserPhysics,pra07}. Here $\gmu$ is the cavity decay rate (of $\hat{a}(t)$) and $W\equiv{2g^2\gper}/(\gper^2+\DLa^2)$ is the stimulated emission rate.
The laser frequency is determined by the line-pulling formula $\gper\DLr + \gmu\DLa=0$, setting a relationship between the atomic detuning $\DLa$ and cavity detuning $\DLr \equiv \oml - \omc$.

A crucial feature of Eq.~(\ref{eq:a1}) is the vanishing of the complex restoring force on $\delta\hat{a}(t)$, i.e. the term proportional to $\delta\hat{a}(t)$ on the right hand side.
This can be attributed to the saturating non-linearity of the laser. The cavity decay as well as the detunings precisely cancel by virtue of the saturating gain and the line-pulling formula.
The photon autocorrelation function is conveniently calculated by expressing $S_{nn}(t) = \langle \dn (t) \dn(0) \rangle$ in terms of the photon number fluctuation operator $\dn (t) = \bar{a}\delta\hat{a}^\dagger(t) + \bar{a}^*\delta\hat{a}(t)$ and solving the coupled Langevin equations for the pair ($\dn, \dD$) (see Appendix \ref{sec:EOM}):
\begin{align}
S_{nn}(\omega) &= \frac{\omega^2+\gpar^2(1+\xi)^2}{(\omega^2-\omega_+^2)(\omega^2-\omega_-^2)} \, WN_g \, \bar{n}.
\label{eq:Snn_aboveTH}
\end{align}
Here $N_g$ is the total number of gain atoms, $\xi\equiv{D_0}/{\bar{D}}-1 = \bar{n}/\nsat$ is the dimensionless saturation factor, $\nsat = \gpar/2W$ is the saturation photon number, and $\omega_\pm = -i\frac{\gpar}{2}\left[(1+\xi)\pm\sqrt{(1+\xi)^2 - 8\xi\frac{\gmu}{\gpar}}\right]$ are the complex relaxation frequencies of the laser. Eq. (\ref{eq:Snn_aboveTH}) shows that the optical damping rate $\Gopt$ vanishes because $S_{nn}(\omega)$ is symmetric with respect to $\omega=0$. This result is valid for $\xi\lesssim1$, $\gpar/\gper \ll 1$ (see Appendix \ref{sec:EOM}) and can be traced back to the vanishing of the complex restoring force on $\delta\hat{a}(t)$.

Within the rate equation approach, the steady-state phonon number can be written as $\bar{n}_m = \bar{n}_m^\text{T} + G^2S_{nn}(-\omM)/\gM$ for $\Gopt=0$ (see Appendix \ref{sec:rateEq}), where $\gM$ is the mechanical damping rate and $\bar{n}_m^\text{T}$ is the phonon number in thermal equilibrium with the mechanical bath alone. The result shows that the mechanical oscillator acts as a spectrometer for the intracavity radiation pressure noise generated by the active medium. It can be shown from \Eq{eq:Snn_aboveTH} that $S_{nn}(-\omM)$ is positive and $\bar{n}_m>\bar{n}_m^\text{T}$, i.e. the optomechanical coupling increases the effective temperature of the mechanical motion, due to the additional noise introduced by the photonic and the atomic medium. The same conclusion can be drawn from the more rigorous approach of integrating the phonon spectrum $S_{b^\dagger b}(\omega)\equiv\int_{-\infty}^\infty \avg{\hat{b}^\dagger(t)\hat{b}(0)}e^{i\omega t} \, dt$~(see Appendix \ref{sec:EOM}).


The situation is vastly different when a seeding signal of frequency $\oml$ is fed into the cavity coherently, in addition to the incoherent pump on the atomic medium. The seeding is described by a time-dependent Hamiltonian $\hat{H}_s = \hb\Oms(\hat{a}^\dagger e^{-i\oml t} + h.c.)$, where $\Oms$ denotes the strength of the seeding signal. Note that $\oml$ of the external laser drive can be chosen differently from the above-threshold laser frequency determined by the line-pulling formula. The system Hamiltonian in the rotating frame of frequency $\oml$ becomes
\begin{align}
\hat{H} &= -\hb[\DLr + G(\hat{b}^\dagger+\hat{b})](\hat{a}^\dagger\hat{a}-\bar{n}) +\hb\omM\hat{b}^\dagger\hat{b} + \frac{\hb\oma}{2}\hat{D}\nonumber \\
&\quad+ \sum_j \hb g(\hat{a}^\dagger\hat{\sigma}^-_j + \hat{\sigma}^+_j\hat{a}) + \hb\Oms(\hat{a}^\dagger + \hat{a}). \label{eq:H2}
\end{align}
The average cavity photon number $\bar{n}$ is now $\Oms$-dependent as well:
\be
\bar{a}\left[1+\frac{g^2\bar{D}}{(i\DLr-\gmu)(\gper-i\DLa)}\right]=\frac{i\Oms}{i\DLr-\gmu}, \label{eq:abar_seed}
\ee
where the steady-state inversion $\bar{D}=D_0/(1+\xi)$ as before. Eq.~(\ref{eq:abar_seed}) can display a bistable behavior \cite{bistability}, but we will focus here to the resolved side band limit ($\gmu\ll|\DLr|$) with a small $\xi\lesssim 1$, where a single steady-state solution exists.

The photon field fluctuation now follows
\be
\dot{\delta\hat{a}}(t) = (i\tDLr-\tilde{\gmu})\delta\hat{a}(t) + \frac{g^2\bar{a} \delta\hat{D}(t)}{\gper-i\DLa} + \Fa(t),
\ee
where the effective detuning $\tDLr$ and cavity decay $\tilde{\gmu}$ are defined as
\be
\tDLr \equiv \DLr + \frac{W \bar{D}}{2\gper}\DLa, \quad \tilde{\gmu} = \gmu - \frac{W\bar{D}}{2}.\label{eq:effective}
\ee
Note that $\bar{D}$ has the same sign as $D_0$, and for $D_0>0$ this leads to an effective cavity decay that is reduced from its intrinsic value.
This tunability of the effective complex frequency of photon fluctuations stems from the fact that $\bar{D}$ is no more clamped at $D_{th}$, as in the case above the laser threshold in the absence of seeding. Instead, it's now a function of the incoherent pump $D_{0}$ and the coherent drive amplitude $|\Oms|$. \Eq{eq:effective} leads to the following form of $S_{nn}$ (see Appendix \ref{sec:EOM}):
\begin{align}
S_{nn}(\omega) &\approx \Lambda(\omega)[(\omega^2+\tDLr^2+\tilde{\gmu}^2)(WN_g+2\gmu)-4\omega\tDLr\tilde{\gmu}], \label{eq:Snn_seeding}
\end{align}
where we have defined $\Lambda(\omega) \equiv \bar{n}\gper^2[\omega^2+\gpar^2(1+\xi)^2]/|\Theta(\omega)|^2$ and $
\Theta(\omega) \equiv \gper[(-i\omega+\tilde{\gmu})^2+\tDLr^2][-i\omega+\gpar(1+\xi)] -2\bar{n}W^2\bar{D}[\DLa\tDLr-\gper(\tilde{\gmu}-i\omega)]$. The previously studied cases can be recovered in various limits of this expression. For the case without seeding $\tilde{\gmu}=\tDLr=0$, $\Theta(\omega)$ becomes $-i\omega\gper(\omega-\omega_+)(\omega-\omega_-)$ and we recover $S_{nn}(\omega)$ in (\ref{eq:Snn_aboveTH}). The passive case is recovered by setting $g = 0$ in Eq.~(\ref{eq:Snn_seeding}), reproducing  the well-known expression $S_{nn}(\omega)=2\gmu\bar{n}/[(\omega+\DLr)^2+\gmu^2]$ \cite{bibnote:notations}.

\begin{figure}[b]
\centering
\includegraphics[width=\linewidth]{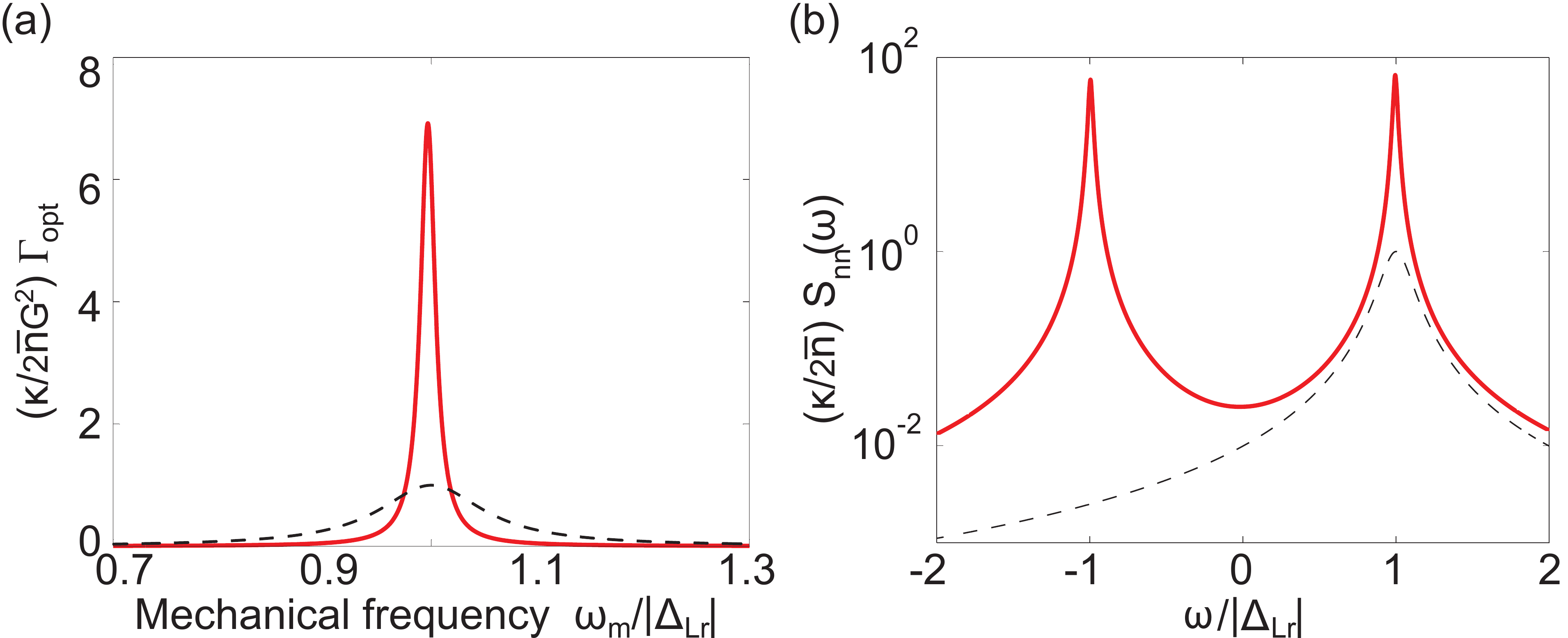}
\caption{(Color online) Optical damping rate (a) and photon autocorrelation function (b) in a seeded laser (solid line) and the corresponding passive system (dashed line). Parameters in this typical example are: $\DLa=-\DLr=1\,\text{GHz}$, $\gper=10\,\text{GHz}$, $\gpar=\gmu=100\,\text{MHz}$, $g=1\,\text{MHz}$, $D_0 = 0.8 N_g = 1.2D_{th}$, and $\bar{n}=10^5$. The resulting $\xi=0.396$, $\gmu/\tilde{\gmu}=7.12$, and $\Gamma_\text{opt,\,max}$ is $6.94$ times its value in the corresponding passive system.
} \label{fig:comparison}
\end{figure}

$S_{nn}(\omega)$ given by (\ref{eq:Snn_seeding}) is asymmetric around $\omega=0$ and leads to a nonvanishing $\Gopt$. For a moderate cavity photon number, $\Theta(\omega)$ is dominated by the first term and
$\Lambda(\omega) \approx \Lambda(-\omega) \approx \bar{n}/[\tilde{\gmu}^2+(\omega-\tDLr)^2][\tilde{\gmu}^2+(\omega+\tDLr)^2]$, indicating that the relaxation frequencies of the seeded laser are roughly $\pm\tDLr-i\tilde{\gmu}$. As a consequence, $S_{nn}(\omega)$ peaks near $\pm\tDLr$ with a width given approximately by $\tilde{\gmu}$ (see Fig.~\ref{fig:comparison}(b)). These peaks are higher than that in the passive case due to the atomic noise, resulting in the optical damping rate
\be
\Gopt(\omega) \approx \frac{-8 G^2 \, \bar{n} \, \omega \tDLr \tilde{\gmu}}{[\tilde{\gmu}^2+(\omega-\tDLr)^2][\tilde{\gmu}^2+(\omega+\tDLr)^2]},\label{eq:GammaOpt2}
\ee
In the resolved side-band limit (of the active case, $\tilde{\gmu} \ll |\tDLr|$), the maximal optical damping is therefore
\be
\Gamma_\text{opt,\,max} \approx \frac{2G^2\bar{n}}{|\tilde{\gmu}|}\left[1+\left(\frac{\tilde{\gmu}}{2\omM}\right)^2\right]^{-1}, \label{eq:GammaOpt_max}
\ee
This results in a gain-enhancement of the damping rate by about $\gmu/\tilde{\gmu}$, when compared to the corresponding passive system with the same mean number of cavity photons (see \Fig{fig:comparison}) \cite{bibnote:nearK0}. The maximal damping rate (\ref{eq:GammaOpt_max}) requires a negative detuning $\tDLr = - \omM$ if the effective cavity decay $\tilde{\gmu}$ is positive, similar to a passive system. If the incoherent pump is strong enough that $\tilde{\gmu}$ becomes negative, a {\it positive} detuning $\tDLr = \omM$ is required instead, otherwise $\Gopt$ becomes negative, resulting in optical heating. We note that the mean number of cavity photons mostly is set by the strength of the coherent drive in the resolved sideband limit (see \Eq{eq:abar_seed}), as in a passive system. Yet the damping rate can be greatly enhanced due to a strongly modified dynamics of the photon number fluctuations.

\begin{figure}[b]
\centering
\includegraphics[width=0.53\linewidth]{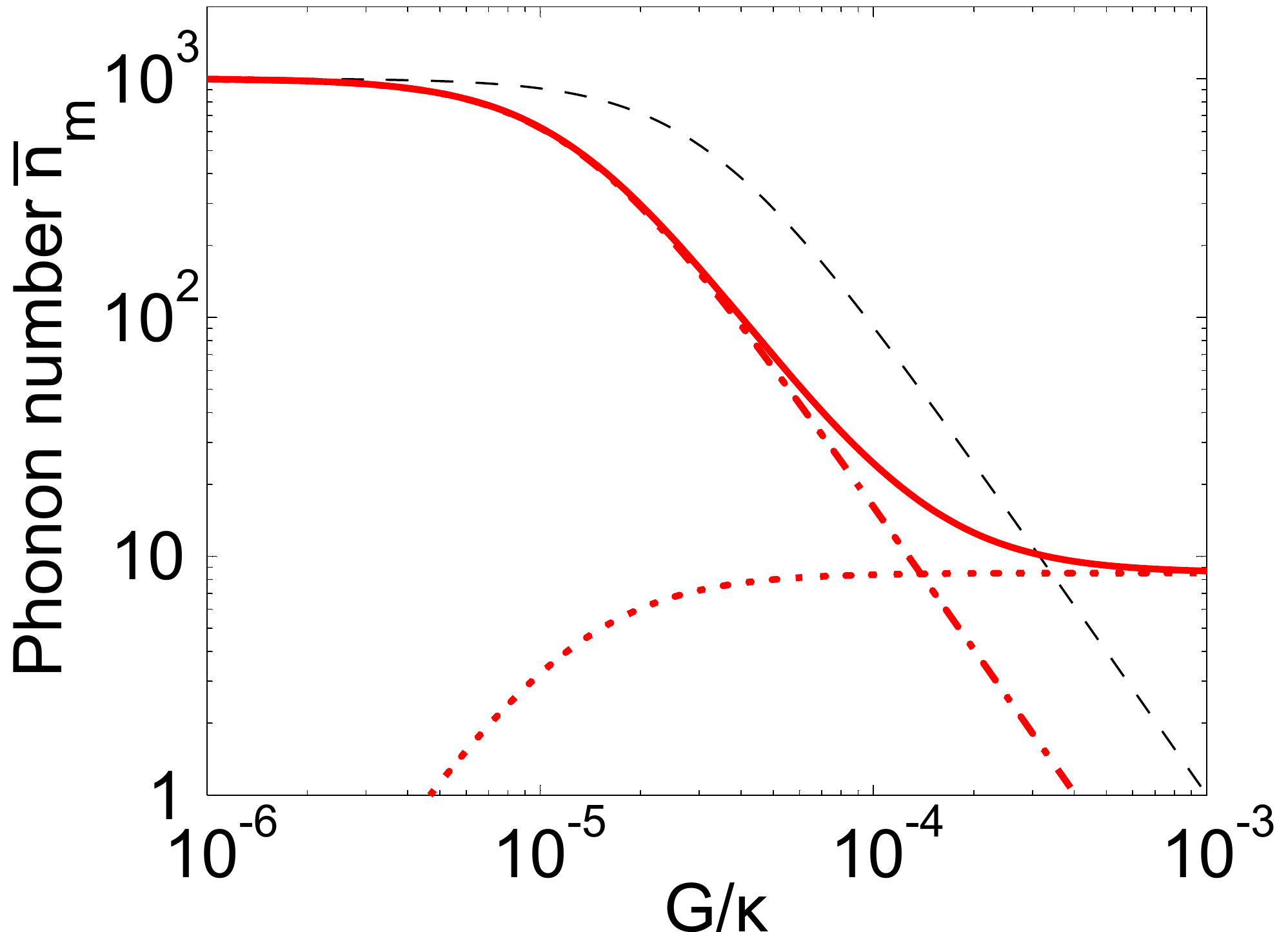}
\caption{(Color online) Steady-state phonon number versus the optomechanical coupling in a seeded laser (solid line) and the corresponding passive system (dashed line). The dotted and dash-dotted lines indicate the contribution from the optical bath and the mechanical bath in the seeded laser, respectively. In the passive system $\bar{n}_m$ comes almost all from the latter. $\bar{n}_m^\text{T}=10^3$, $\omM=-\DLr$, $\Gamma_m/\gmu=2\times10^{-4}$, and the other parameters are the same as in Fig.~\ref{fig:comparison}.} \label{fig:changeG}
\end{figure}

We note that the enhancement of $\Gopt$ does not necessarily lead to a reduced temperature of the effective optical bath, which can be represented by $\bar{n}_m^\text{opt} = (S_{nn}(\omM)/S_{nn}(-\omM) - 1)^{-1}$ (see Appendix \ref{sec:rateEq}). The total phonon number is given by $\bar{n}_m = (\Gopt\bar{n}_m^\text{opt} + \gM\bar{n}_m^\text{T})/(\Gopt+\gM)$, and $\bar{n}_m^\text{opt}(<\bar{n}_m^\text{T})$ is the minimum phonon number, attained in the limit $\Gopt\gg\gM,\gM\bar{n}_m^\text{T}/\bar{n}_m^\text{opt}$. In this limiting regime the addition of gain leads to a higher effective temperature of the mechanical motion, since $\bar{n}_m^\text{opt}$ is increased due to the enhancement of both $S_{nn}(\omM)$, $S_{nn}(-\omM)$ by the atomic noise:
\be
\bar{n}_m^\text{opt} \approx\frac{WN_g+2(\gmu-\tilde{\gmu})}{4\tilde{\gmu}}.\label{eq:nApprox}
\ee
However in the regime $\gM\ll\Gopt\ll\gM\bar{n}_m^\text{T}/\bar{n}_m^\text{opt}$ that is common in passive systems, either due to a small $\bar{n}_m^\text{opt}$, a relatively high thermal temperature, or a small $G$, the thermal contribution $\gM\bar{n}_m^\text{T}/\Gopt$ in $\bar{n}_m$ cannot be neglected in general. In this parameter regime we find that the cooling of the mechanical motion can be more effective with the addition of the gain, due to the stronger suppression of the thermal contribution as a consequence of an enhanced $\Gopt$.

\begin{figure}[tpb]
\centering
\includegraphics[width=\linewidth]{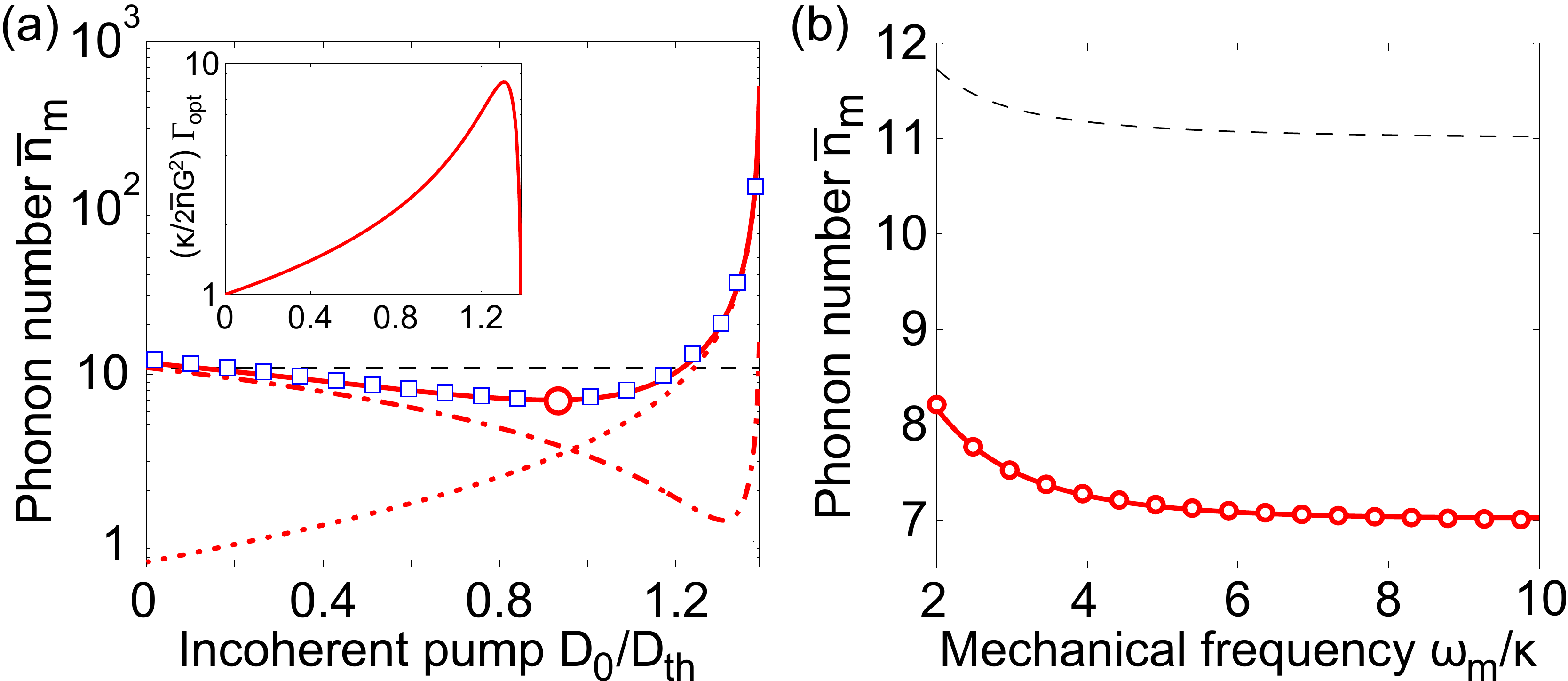}
\caption{(Color online) (a) Steady-state phonon number versus incoherent pump strength in a seeded laser (solid line). The dotted and dash-dotted lines indicate the contribution from the optical bath and the mechanical bath, respectively.
The circle and squares show the approximation (\ref{eq:nApprox2a}),(\ref{eq:nApprox2b}) and the result from integrating the phonon spectrum, respectively. The dashed line indicates the phonon number in the corresponding passive system. $G/\gmu=3\times10^{-4}$, $\bar{n}_m^\text{T}=10^3$, and the other parameters are the same as in Fig.~\ref{fig:changeG}. $\tilde{\gmu}$ becomes zero at $D_0/D_{th}=(1+\xi)\simeq1.396$. Inset: Optical damping rate in the seeded laser \cite{bibnote:nearK0}. (b) Minimized phonon number at different mechanical frequencies via the optimization of $D_0$. $\DLr=-\omM$ is adjusted accordingly. Dots show the fitting using \Eq{eq:nApprox2b} multiplied by $[1+6(\tilde{\gmu}/\omM)^2]$.} \label{fig:nPhonon_D0}
\end{figure}

One such example is given in Fig.~\ref{fig:changeG}, in which $\bar{n}_m^\text{opt}$ is fixed and the weight of the optical and thermal contributions in $\bar{n}_m$ is adjusted by considering different values of $G$. For $G/\gmu\lesssim 3\times10^{-4}$ the seeded laser studied in Fig.~\ref{fig:comparison} has a lower phonon number, which has a minimum close to
$\bar{n}_m^\text{opt}\simeq9$. We note that the upper limit of this range scales linearly with $(\bar{n}_m^\text{T}/\bar{n})^{\frac{1}{2}}$.

For a fixed $G$ and $\bar{n}$, the incoherent pump strength $D_0$ provides a means to minimize $\bar{n}_m$. If the optical or thermal contribution in $\bar{n}_m$ is dominant for {\it all possible values} of $D_0$, the optimization is straightforward and achieved with the lowest $\bar{n}_m^\text{opt}$ or the largest enhancement of $\Gopt$, respectively. When this is not the case, the optimization is achieved when $\bar{n}_m^\text{opt}$ roughly equals the thermal contribution, i.e.
\be
\tilde{\gmu} \approx \left[\frac{G^2\bar{n}(WN_g+2\gmu)}{2\Gamma_m\bar{n}_m^\text{T}}\right]^{\frac{1}{2}},\label{eq:nApprox2a}
\ee
at which the phonon number is approximately their geometric mean
\be
\bar{n}_m \approx \left[\frac{(WN_g+2\gmu)\Gamma_m\bar{n}_m^\text{T}}{2G^2\bar{n}}\right]^{\frac{1}{2}}-\frac{1}{2}.\label{eq:nApprox2b}
\ee
We have taken $\omM=-\DLr$ and neglected the weak $\omM$-dependence in (\ref{eq:nApprox2a}) and (\ref{eq:nApprox2b}), which is only second order in $\tilde{\gmu}/\omM$ as illustrated in \Fig{fig:nPhonon_D0}(b). This implies that the optimization of $D_0$ is achieved almost simultaneously for all the mechanical frequencies in the resolved sideband limit.

\begin{figure}[t]
\centering
\includegraphics[width=0.6\linewidth]{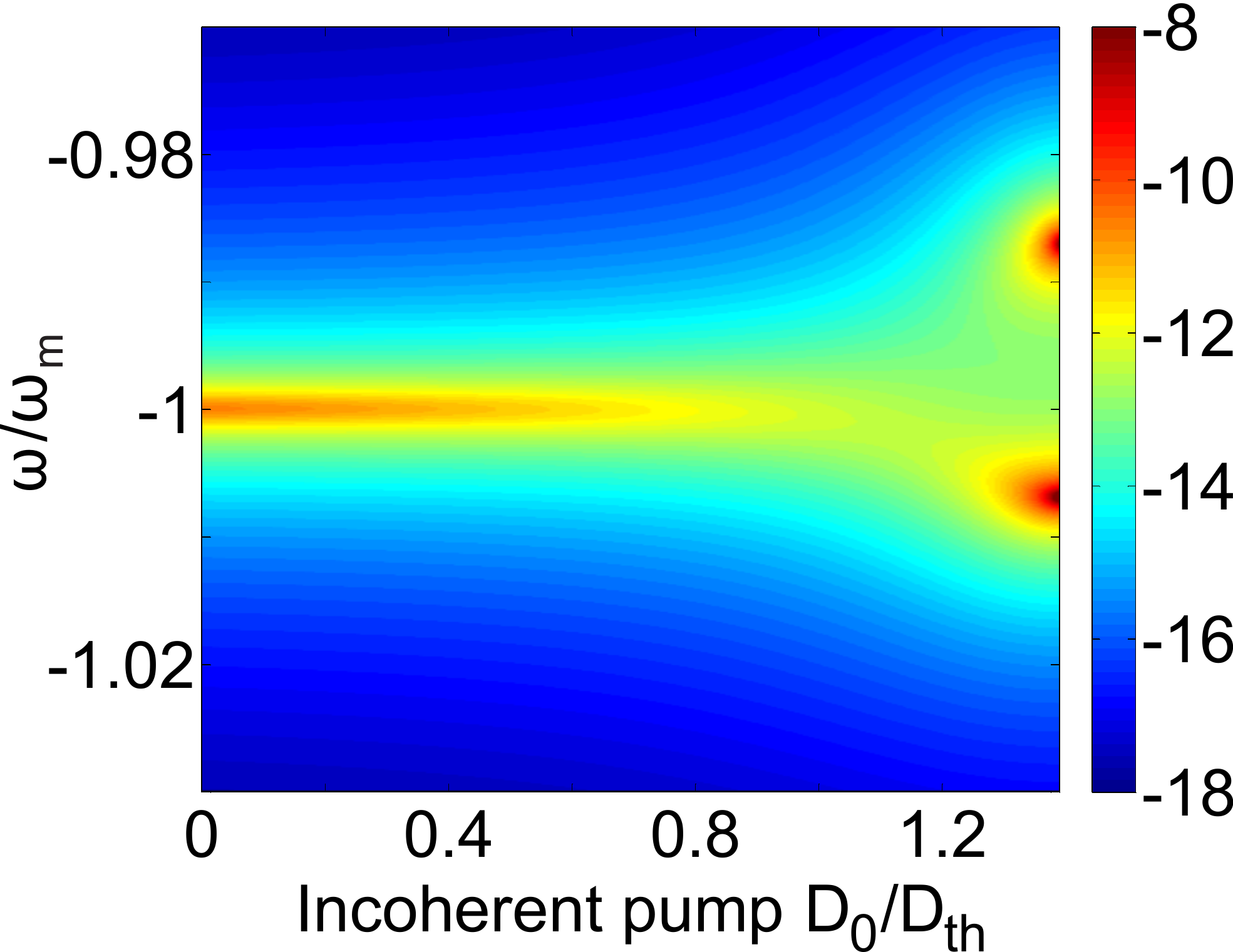}
\caption{(Color online) Phonon spectrum $\log_{10}(S_{b^\dagger b})$ for the seeded laser in \Fig{fig:nPhonon_D0}(a), showing optomechanical normal mode splitting as a function of the incoherent pump strength. The system is in the weak optomechanical coupling regime in the absence of the gain (i.e. $D_0=0$).} \label{fig:Sbb}
\end{figure}

$\bar{n}_m$ calculated by the rate equation agrees well with results obtained by integrating the phonon spectrum, i.~e.  $\bar{n}_m = \frac{1}{2\pi}\int d\omega\,\avg{\hat{b}^\dagger(\omega)\hat{b}(-\omega)}$, as shown in \Fig{fig:nPhonon_D0}(a). We note that $\bar{a}G\simeq 0.01\omM$ in this example, and it is more than one order of magnitude smaller than $\tilde{\gmu}\gM$ for a small $D_0$. This places the system in the weak optomechanical coupling regime, and in the phonon spectrum there is only one peak near $\omega=-\omM$ (see \Fig{fig:Sbb}). The reduction of $\bar{n}_m$ as $D_0$ increases from 0 results in the lowering and a slight broadening of this peak \cite{bibnote:2ndpeak}. As $D_0$ is further increased, $\tilde{\gmu}\rightarrow 0$ and the system undergoes a transition to the strong optomechanical coupling regime, since $\tilde{\gmu}\gM$ becomes comparable and even smaller than $\bar{a}G$, which is kept fixed in this process. Consequently, the peak at $\omega=-\omM$ in the phonon spectrum splits into two, resulting in hybrid resonances separated by roughly $2\bar{a}G$ \cite{Marquardt07}; their enhancement with $D_0$ reflects the increase of $\bar{n}_m$ in this regime.

In summary, we have discussed the optical cooling of a cavity mirror in the presence of an incoherently pumped atomic medium. In the absence of an external laser drive, the coherent radiation in the cavity generated by the incoherent pump above threshold results in a vanishingly small optical damping rate. This always results in the heating of the cavity mechanical oscillator due to the additional photon noise generated by the atomic medium. In the presence of an additional external coherent seeding, we show that the cooling rate and the minimum attainable phonon number can be tuned through the strength of the incoherent pump, which can be enhanced and lowered respectively from their passive cooling values. In addition, we showed that the strong optomechanical coupling regime can be reached by increasing the incoherent pump strength, even at a fixed number of cavity photons.

We acknowledge helpful discussions with Girish Agarwal. L.~G. and H.~E.~T. acknowledge NSF for Grant MIRTHE EEC-0540832. F.~M. acknowledges DARPA ORCHID, an ERC Starting Grant, DFG Emmy-Noether, and the ITN cQOM.

\appendix
\section{Equations of motion}
\label{sec:EOM}

The optical field in the cavity couples to the mechanical motion of its mirror and the polarization of the optically active medium with strengths $G\equiv\frac{x_{\text{ZPF}}}{L}\omc$ and $g=-\mu \sqrt{\omc/\epso V}$, respectively:
\be
\dot{\hat{a}} = (-i\omc-{\gmu})\hat{a} - ig\hat{P} + iG(\hat{b}^\dagger+\hat{b})\hat{a} + \hat{\mathcal{F}}_{\gmu}. \label{eq:a}
\ee
$\gmu$ is the cavity decay rate, $\mu$ is the dipole strength, $\epso$ is the permittivity of the vacuum and $V$ is the effective volume. $\mathcal{F}$'s here and below are the fluctuating Langevin forces associated with the corresponding decay channels. In the rate equation approach the mechanical backaction on the dynamics of the rest of the system is neglected, which we will restore in the discussion of the phonon spectrum.

The polarization $\hat{P}$ and inversion $\hat{D}$ follow
\begin{align}
&\dot{\hat{P}} = (-i\oma-\gper)\hat{P} + ig\hat{D}\hat{a} + \hat{\mathcal{F}}_{\gper}, \label{eq:sigma-}\\
&\dot{ \hat{D}} = \gpar(D_0-\hat{D}) + 2ig(\hat{a}^\dagger \hat{P} - \hat{P}^\dagger\hat{a}) + \hat{\mathcal{F}}_{\gpar}. \label{eq:d}
\end{align}
The incoherent pump strength is characterized by the parameter $D_0$ which is the steady-state value of the inversion created in the absence of a cavity field, and $\gper,\,\gpar$ are the decay rate of the polarization and inversion, respectively.
Assuming $\gpar\ll\gper$ as in a Class A or B laser, we adiabatically eliminate $\hat{P}$ in the rotating frame of the laser frequency $\oml$ \cite{Haken}:
\be
\hat{P} \approx \frac{ig\hat{a}\hat{D} + \hat{\mathcal{F}}_{\gper}}{\gper - i\DLa}. \label{eq:sigma-approx}
\ee
The linearization of the operators $\hat{a},\, \hat{a}^\dagger,\,\hat{D}$ about their steady-state value $\bar{a}$, $\bar{a}^*$, $\bar{D}$ in the same rotating frame leads to \begin{align}
\dot{\delta\hat{a}} = (i\DLr-{\gmu}) \delta\hat{a} + \frac{g^2[\bar{D} \delta\hat{a} + \bar{a} \delta\hat{D}]}{\gper-i\DLa}  + \Fa ,\label{eq:a0}
\end{align}
where the effective fluctuation force on $\delta\hat{a}$ is given by
\be
\Fa = \hat{\mathcal{F}}_{\gmu} - \frac{ig}{\gper - i\DLa} \hat{\mathcal{F}}_{\gper}.
\ee
$\Fa$ satisfies $\langle \Fa (t) \Fa^\dagger (t') \rangle = \Daad \delta(t-t')$, $\langle \Fa^\dagger(t) \Fa (t')\rangle = \Dada \delta(t-t')$, where the diffusion coefficients $\Daad = \Daad^\text{T} + \Daad^\text{SE}$ and $\Dada=\Dada^\text{T} + \Dada^\text{SE}$ contain a blackbody contribution $\Daad^\text{T} = 2\gmu (\bar{n}^\text{T}+1)$, $\Dada^\text{T} = 2\gmu \bar{n}^\text{T}$ and a contribution due to collective atomic dissipation processes
\begin{align}
&\Dada^\text{SE} = \frac{W}{2}\left[(N_g+\bar{D})+\frac{\gpar}{2\gper}(D_0-\bar{D})\right],\label{eq:D_eed}\\
&\Daad^\text{SE} = \frac{W}{2}\left[(N_g-\bar{D})-\frac{\gpar}{2\gper}(D_0-\bar{D})\right].\label{eq:D_ede}
\end{align}
$\bar{n}^\text{T}$ is the number of blackbody photons inside the cavity, vanishingly small at optical frequencies studied here and $N_g$ is the total number of gain atoms.

Eq.~(\ref{eq:a0}) is further simplified to (4) in the main text by using the values of $\bar{D}$ and $\bar{a}$. Similarly, we find
\be
\dot{\delta\hat{D}} = -(\gpar+2W\bar{n})\delta\hat{D} - 2W\bar{D}[\bar{a}\delta\hat{a}^\dagger + \bar{a}^*\delta\hat{a}] + \Fd, \label{eq:d1}
\ee
where $\bar{n}$ is the average cavity photon number and $W\equiv{2g^2\gper}/(\gper^2+\DLa^2)$ is the stimulated emission rate. The fluctuation force $\Fd$
satisfies $\avg{\Fd\Fd}=\Ddd\delta(t-t')$, $\avg{\Fa\Fd}=\Dad\delta(t-t')$, $\avg{\Fa^\dagger\Fd}=\Dadd\delta(t-t')$ \cite{Haken}, where
\begin{align}
\Ddd &= 2\gpar\left(N_g-\frac{D_0\bar{D}}{N_g}\right), \\
\Dad &= \frac{\gpar g^2\bar{D}\bar{a}}{(\gper-i\DLa)^2}\left(1-\frac{D_0}{N_g}\right),\\
\Dadd &= -\frac{\gpar g^2\bar{D}\bar{a}}{(\gper+i\DLa)^2}\left(1+\frac{D_0}{N_g}\right).
\end{align}
The equations of motion for $\delta\hat{n}(t) = \bar{a}\delta\hat{a}^\dagger(t) + \bar{a}^*\delta\hat{a}(t)$ and $\delta\hat{D}(t)$ in the frequency domain can then be written as
\be
\label{eq:linearResponse}
\mathcal{M}_2
\begin{pmatrix}
\delta \hat{n}(\omega) \\
\delta \hat{D}(\omega)
\end{pmatrix} =
\begin{pmatrix}
\bar{a}\Fa^\dagger (\omega) + \bar{a}^*\Fa (\omega) \\
\Fd
\end{pmatrix},
\ee
where
\be
\mathcal{M}_2 \equiv
\begin{pmatrix}
-i\omega & {-\xi\gpar/2} \\
2W\bar{D} & -i\omega + \gpar(1+\xi)
\end{pmatrix}.
\label{eq:M2}
\ee
Here $\xi\equiv{D_0}/{\bar{D}}-1 = \bar{n}/\nsat$ is the dimensionless saturation factor, which is a convenient parameter used in the main text to measure the number of photons generated, and $\nsat = \gpar/2W$ is the saturation photon number. Solving (\ref{eq:linearResponse}) we find
\begin{align}
\delta n(\omega) &= \frac{i\omega-\gpar(1+\xi)}{(\omega-\omega_+)(\omega-\omega_-)}[\bar{a} \Fa^\dagger (\omega) + \bar{a}^* \Fa (\omega)] \nonumber \\
&\quad-\frac{\xi\gpar}{2(\omega-\omega_+)(\omega-\omega_-)}\Fd,\label{eq:delta_n}
\end{align}
which lead to
\begin{align}
S_{nn}(\omega) &= \frac{\omega^2+\gpar^2(1+\xi)^2}{(\omega^2-\omega_+^2)(\omega^2-\omega_-^2)} \, \bar{n} \, ( \Daad + \Dada ) \nonumber \\
&+\frac{\gpar^2\xi^2}{4(\omega^2-\omega_+^2)(\omega^2-\omega_-^2)} \Ddd \nonumber \\
&+\frac{(1+\xi)\xi\gpar^2\bar{a}}{(\omega^2-\omega_+^2)(\omega^2-\omega_-^2)}\re[\Dad+\Dadd]  \nonumber \\
&+\frac{\omega\xi\gpar\bar{a}}{(\omega^2-\omega_+^2)(\omega^2-\omega_-^2)}\im[\Dad+\Dadd]  \label{eq:Snn_aboveTH3}
\end{align}
using the Wiener-Khinchin theorem $S_{nn} (\omega) = \langle \dn (\omega) \dn (-\omega) \rangle$. We have set  $\bar{a}=\bar{a}^*$ by choosing a proper $t=0$. $\omega_\pm$ are the complex relaxation frequencies of the laser as defined in the main text.

We first note that the first terms in \Eq{eq:Snn_aboveTH3} is much larger than the rest for $\xi\lesssim1$, $\gpar/\gper \ll 1$, with which Eq.~(5) in the main text is derived. Furthermore, we have neglected the blackbody contribution to the noise, i.e. taking $\Dada+\Daad = WN_{g} + 2\gmu(2\bar{n}^\text{T}+1)\approx WN_g$ in the first term. It is symmetric with respect to $\omega=0$, and so are the second and third terms in \Eq{eq:Snn_aboveTH3}. The asymmetry of $S_{nn}(\omega)$ only arises from the last term and is vanishingly small. This derivation shows that $\Fd$ only leads to a small effect, and we have thus neglected it in the rest of the discussions in the main text.

\begin{figure}[h]
\centering
\includegraphics[width=.6\linewidth]{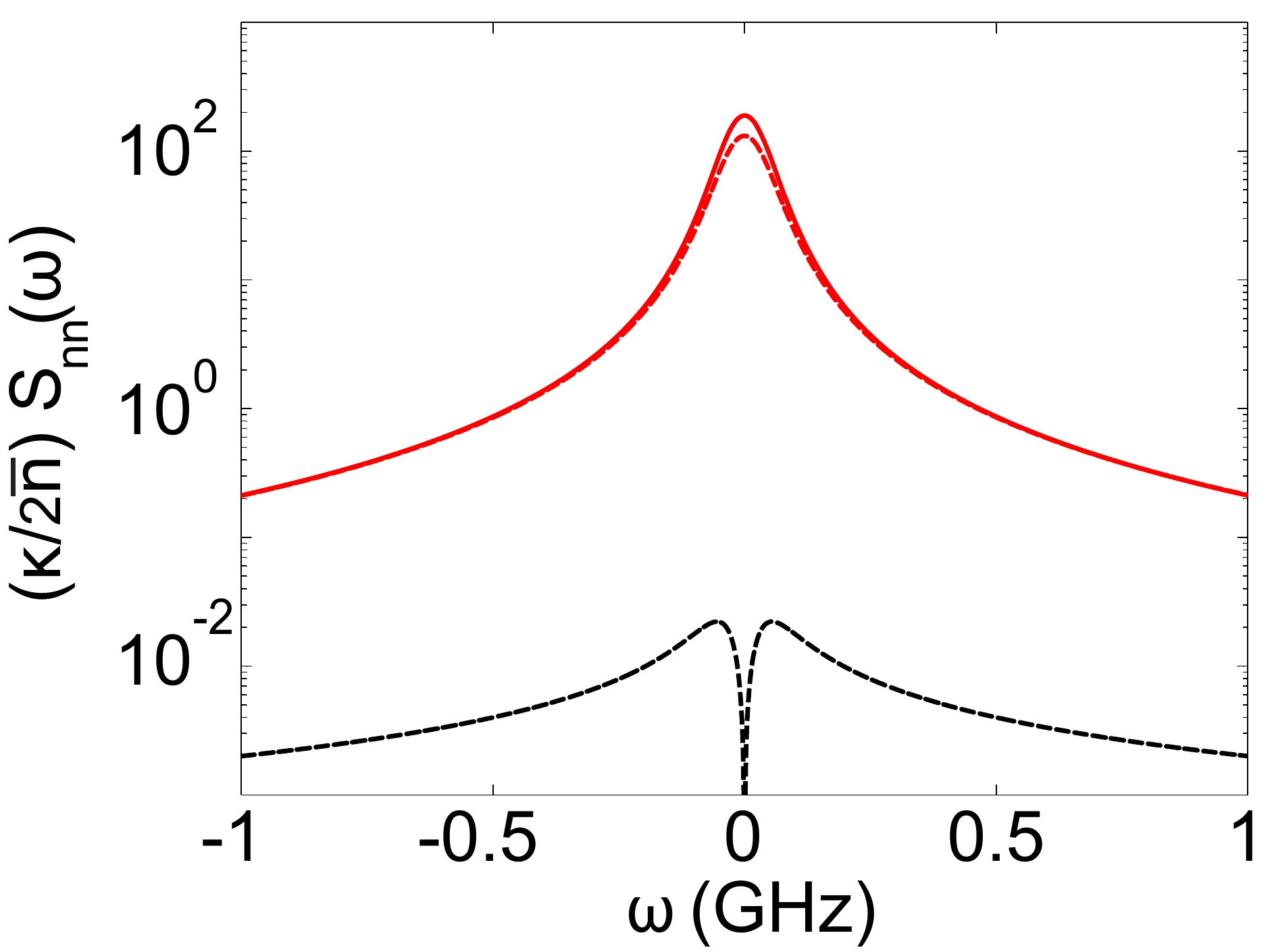}
\caption{ Photon autocorrelation function calculated with the polarization $\hat{P}(t)$ adiabatically eliminated (solid) and linearized (dashed). They agree well, with the asymmetry of the latter (i.e. $|S_{nn}(\omega)-S_{nn}(-\omega)|$) shown by the dash-dotted line. Parameters used are: $\gper=10\,\text{GHz}$, $\gpar=\gmu=100\,\text{MHz}$, $\DLa=10$~GHz, $g=1\,\text{MHz}$, $\bar{n}=10^5$, and $D_{th}=0.1N_g$.
} \label{fig:linearP}
\end{figure}

The main approximation we have employed in deriving Eq.~(5) in the main text is the adiabatic elimination of $\hat{P}(t)$ (see \Eq{eq:sigma-approx}). This is typically a good approximation for a Class A or B laser. To demonstrate its validity in the calculation of $\Gopt$, below we compare our results to the solution of the full linearized equations of motion for the photonic and atomic degrees of freedom, i.e. keeping the fluctuation $\delta\hat{P}(t)$ defined by $\hat{P}(t)=(\bar{P}+\delta\hat{P}(t))e^{-i\oml t}$. $\delta\hat{P}(t)$ evolves according to
\be
\dot{\delta\hat{P}} = (i\DLa-\gper)\delta\hat{P} + ig(\bar{D}\delta\hat{a} + \bar{a}\delta\hat{D}) + \F_{\gper}.
\ee
By solving it together with its hermitian conjugate and $\delta\hat{a},\,\delta\hat{a}^\dagger, \delta\hat{D}$, we again derive an expression for $S_{nn}(\omega)$. Due to its complexity, here we only compare it with Eq.~(5) numerically. As Fig.~\ref{fig:linearP} shows, the asymmetry of $S_{nn}(\omega)$ about $\omega=0$ arising from the fast dynamics of $\delta\hat{P}(t)$ is extremely small, so is the optical damping rate $\Gopt$.

The vanishing of the complex frequency of $\delta\hat{a}^\dagger(t)$ Eq.~(4) in the main text decouples the dynamics of $\delta \hat{a}$ and $\delta \hat{a}^\dagger$ for the unseeded case. The situation is different with a non-vanishing seeding signal, where the phase is enslaved by the phase of the coherent driving field. We therefore introduce $\delta \hat{u}(t)\equiv\bar{a}^*\delta\hat{a}(t) - \bar{a}\delta\hat{a}^\dagger(t)$ along with $\delta \hat{n}(t)$ and $\delta \hat{D}(t)$. Their equations of motion with the Hamiltonian (6) in the main text can be written as
\be
\mathcal{M}_3
\begin{pmatrix}
\delta \hat{n}(\omega) \\
\delta \hat{u}(\omega) \\
\delta \hat{D}(\omega)
\end{pmatrix} =
\begin{pmatrix}
\bar{a}^* \Fa (\omega) + \bar{a} \Fa^\dagger(\omega) \\
\bar{a}^* \Fa (\omega) - \bar{a} \Fa^\dagger(\omega) \\
0
\end{pmatrix},
\ee
where
\be
\mathcal{M}_3 \equiv
\begin{pmatrix}
-i\omega + \tilde{\gmu} & -i\tilde{\Delta}_{Lr} & {-W\bar{n}} \\
-i\tilde{\Delta}_{Lr} & -i\omega + \tilde{\gmu} & \frac{-iW\bar{n}\tilde{\Delta}_{Lr}}{\gper} \\
2W\bar{D} & 0 & -i\omega + \gpar(1+\xi)
\end{pmatrix}.\label{eq:linearResponse2}
\ee
Their solution leads to the autocorrelation function (10) in the main text. To recover \Eq{eq:linearResponse} in the unseeded case, we note that $\delta\hat{u}$ is decoupled from $\delta\hat{n}$ and $\delta\hat{D}$ since $\tilde{\gmu}=\tilde{\Delta}_{Lr}=0$.

Next we discuss the calculation of the phonon spectrum $S_{b^\dagger b}=\avg{\hat{b}^\dagger(\omega)\hat{b}(-\omega)}$ with the linearized equations of motion. The dynamical equation for the mechanical motion is
\be
\dot{\hat{b}} = (-i\omM-\frac{\gM}{2})\hat{b} + iG(\hat{a}^\dagger\hat{a}-\bar{n}) + \hat{\mathcal{F}}_{\gM}, \label{eq:b}
\ee
where $\omM$ and $\gM$ are the frequency and the intrinsic damping rate of the mechanical oscillator. Solving it together with $\delta \hat{a}(t),\,\delta \hat{a}^\dagger(t)$ (or $\delta\hat{n}(t)$) and $\delta \hat{D}(t)$ in the frequency domain, we find in general
\be
\hat{b}(\omega) = \chi_a(\omega)\F_a + \chi_{a^\dagger}(\omega)\F_a^\dagger + \chi_b(\omega)\F_{\gM} + \chi_{b^\dagger}(\omega)\F_{\gM}^\dagger.\label{eq:bw}
\ee
For an unseeded laser, we find that $\chi_{b^\dagger}=0$, $\chi_m(\omega)=[i(\omM-\omega)+\gM/2]^{-1}$ are exactly the same as in the case of a standalone mechanical oscillator, as a consequence of the vanishing of the complex restoring force on $\delta\hat{a}(t)$.  The latter also leads to $\chi_a(\omega)=\chi_{a^\dagger}(\omega)$, given by
\be
\chi_a(\omega)=-\frac{\bar{a}G[\omega+i\gpar(1+\beta)]}{(\omega-\omega_+)(\omega-\omega_-)}\chi_m(\omega).
\ee
The resulting phonon number $\bar{n}_m = \frac{1}{2\pi}\int_{-\infty}^{\infty} S_{b^\dagger b}(\omega)d\omega$ is then the sum of the unaltered thermal contribution from the mechanical bath,
\be
\bar{n}_m^\text{T}= \frac{1}{2\pi}\int_{-\infty}^{\infty}|\chi_m(-\omega)|^2 \mathcal{D}_{b^\dagger b}\,d\omega = \frac{\mathcal{D}_{b^\dagger b}}{\gM},
\ee
and the extra contribution form the optical noise $\frac{1}{2\pi}\int_{-\infty}^{\infty} \,d\omega \, |\chi_a(-\omega)|^2 [WN_{g} + 2\gmu(2\bar{n}^\text{T}+1)]>0$.

In a seeded laser the response functions are more complicated, and here we only give $\chi_a(\omega)$ as an example
\begin{widetext}
\be
\label{eq:chi_seeded}
\chi_a(\omega)=-\frac{i\bar{a}G\chi_r^*(-\omega)\chi_m(\omega)}
{1+4\bar{n}G^2\omM\tDLr\chi_r^*(-\omega)\chi_r(\omega)\chi_m^*(-\omega)\chi_m(\omega) + \frac{2i\bar{n}W\bar{D}g^2}{\omega+i(1+\xi)\gpar}\left[\frac{\chi_r^*(-\omega)}{\gper-i\Delta}+\frac{\chi_r(\omega)}{\gper+i\Delta}\right]},
\ee
\end{widetext}
where $\chi_r(\omega)=[i(\tDLr-\omega)+\tilde{\gmu}]^{-1}$. The transition of the system from the weak optomechanical coupling regime to the strong optomechanical coupling regime can be analyzed from the denominator of \Eq{eq:chi_seeded}. We first neglect the last term, which is absent in the passive system. The hybrid resonances appear when the real part of the denominator becomes zero, which occurs at $\omega\approx-\omM\pm\bar{a}G$ and requires that $\bar{a}G$ is comparable or larger than $\tilde{\gmu}\gM$. The last term and the minute difference between $\tDLr$ and $\DLr$ slightly change the separation of the hybrid resonances, as well as making them asymmetric about $\omega=-\omM$.

\section{Rate equation model for mechanical oscillation}
\label{sec:rateEq}
In the main text we have used a rate equation approach, where the rates are calculated in the quantum noise picture as in Ref.~\cite{Marquardt07}. This approach is a good approximation if the relaxation of the effective optical bath (e.g. $\tilde{\gmu}$ in a seeded laser) is much faster than its energy exchange rate with the mechanical oscillator (i.e. $\Gopt$), with which we can treat the former as a Markovian bath. This condition is satisfied for $D_0/D_{th}\lesssim1$ in Fig.~3(a), and its prediction is qualitatively correct even beyond this range as we find.

Denoting $\rho_{n,n}$ as the occupation probability in the state with $n$ phonons and $\Gamma_{n,n\pm1}(\omM)$ as the transition rate from the state with $n$ phonons to the state with $n\pm1$ phonons, the detailed balance in equilibrium requires
\be
\rho_{n,n}\Gamma_{n,n-1}(\omM)+\rho_{n-1,n-1}\Gamma_{n-1,n}(\omM) = 0,
\ee
i.e. the transition from the state with $n$ phonons to the state with ($n-1$) phonons are balanced by the reversed transition. Using the system-bath theory or Fermi's golden rule, we find that $\Gamma_{n,n-1}$ and $\Gamma_{n-1,n}$ are proportional to $n$, and we denote $\Gamma_{1,0}\equiv\Gamma_\downarrow$ and $\Gamma_{0,1}\equiv\Gamma_\uparrow$.

Assuming the equilibrium satisfies a thermal distribution, one can easily show that the average phonon number
\be
\bar{n}_m = \frac{\Gamma_\uparrow}{\Gamma_\downarrow-\Gamma_\uparrow}. \label{eq:rate1}
\ee
If the transitions are only caused by the intrinsic mechanical damping, the average phonon number is just the thermal phonon number $\bar{n}_m^\text{T}$. Similarly, we can define a $\bar{n}_m^\text{opt}$ if the transitions are only caused by the optical damping, and its effective temperature is give by
\be
T_\text{opt} = \frac{\hbar\omM}{k_B}\left[\ln\avg{\bar{n}_m^\text{opt}+1}-\ln\avg{\bar{n}_m^\text{opt}}\right],
\ee
where $k_B$ is the Boltzmann constant. With both damping processes present, $\bar{n}_m$ can be rewritten as a weighted average of these two phonon numbers
\be
\bar{n}_m = \frac{\Gopt\bar{n}_m^\text{opt} + \gM\bar{n}_m^\text{T}}{\Gopt+\gM}. \label{eq:rate2}
\ee

$\bar{n}_m^\text{opt}$ becomes ill defined if $\Gopt$ vanishes, as in an unseeded laser. In this case $\bar{n}_m$ can be expressed as
\be
\bar{n}_m = \frac{\Gamma^\text{opt}_\uparrow}{\gM} + \bar{n}_m^\text{T} = \frac{G^2S_{nn}(-\omM)}{\gM} + \bar{n}_m^\text{T}. \label{eq:rate3}
\ee
It has the same structure as the result from integrating the phonon spectrum $S_{b^\dagger b}$, i.e. the unaltered thermal photon number plus a contribution from the photon noises. We note that here the relaxation rate of the optical bath is not $\tilde{\gmu}(\approx0)$ but rather $-\im[\omega_\pm]$($\neq0$), and the rate equation still applies.


\begin{thebibliography}{99}
\bibitem{bibnote:notations} Note that the cavity decay rate ``$\gmu$'' in Ref.~\cite{Marquardt07} is expressed in terms of the photon number, and it is $2\gmu$ in our current notation.
\bibitem{bibnote:2ndpeak} The other peak of the phonon spectrum near $\omega=\omM$ can be neglected in the discussion of the phonon number here. This is because it is minute comparable with that near $\omega=-\omM$ if $\omM=-\DLr$; their ratio at $D_0=0$ is $\gM\gmu/2\omM^2$, which is $10^{-6}$ in the case shown in \Fig{fig:Sbb}.
\end{thebibliography}
\end{document}